# Predictive Modeling of Rat Brain Local Field Potentials using Single-Variable and Multivariable Approaches

AmirAli Kalbasi [a], Shole Jamali [b], Mahdi Aliyari Shoorehdeli [a, *], Abbas Haghparast[c]

[a] *Department of Mechatronics, Faculty of Electrical Engineering, K. N. Toosi University of Technology, Tehran, Iran*

[b] *Department of Neuroscience, Medical University of South Carolina, Charleston, SC, USA*

[c] *Neuroscience Research Center, School of Medicine, Shahid Beheshti University of Medical Sciences, Tehran, Iran*

## Abstract

Accurate prediction of neural dynamics in the brain's reward circuitry is crucial for elucidating how natural and pharmacological rewards influence neural activity and connectivity. Traditional linear models, such as autoregressive (AR) and vector autoregressive (VAR), often inadequately capture the inherent nonlinear interactions in neural data. This study develops and benchmarks both linear and advanced deep learning models for predicting local field potentials (LFPs) in the rat hippocampus (HIP) and nucleus accumbens (NAc) across morphine, food, and saline conditions. We compared AR, VAR, long short-term memory (LSTM), and wavelet-based deep learning model (WCLSA). Additionally, a novel wavelet coherence-enhanced model (WCOH CLSA) was introduced to capture cross-region connectivity. Results indicate that WCLSA achieves superior predictive accuracy (up to 0.97 for HIP in food, 0.96 for NAc in morphine), while VAR performs competitively in the food group due to significant HIP-NAc correlation. Wavelet coherence analysis reveals robust connectivity in natural reward contexts and disrupted or nonlinear relationships under pharmacological influence. These findings highlight the differential engagement of HIP and NAc in reward processing and underscore the importance of advanced nonlinear models for capturing complex neural dynamics. The study provides a robust framework for predictive neuroscience and elucidates functional interactions within the reward circuitry.

## 1. Introduction

Predicting neural dynamics is a powerful method for understanding brain activity, guiding clinical interventions, and designing brain–computer interfaces. Prediction models can anticipate future states while uncovering patterns and mechanisms in complex systems like the reward circuitry. By capturing the temporal evolution of neural signals, these models forecast responses to diverse stimuli and support real-time interventions. They help predict how neural activity evolves under various conditions, such as natural versus pharmacological rewards, revealing the interplay between brain regions. This capability is vital for targeted interventions—like closed-loop neuromodulation systems—that rely on real-time predictions to modulate activity therapeutically [1]. Accurate predictions also enhance our understanding of neural circuits, aiding the design of precise interfaces. The reward circuitry, especially the Hippocampus (HIP) and Nucleus Accumbens (NAc), is critical in processing both natural (e.g., food) and drug (e.g., morphine) rewards, yet their distinct roles and dynamic interplay remain unclear. Developing accurate Local Field Potentials (LFPs) prediction models for these regions is essential. This study examines these dynamics across morphine, food, and saline groups to assess how rewards modulate neural activity and connectivity [2].

Traditional forecasting methods for neurophysiological time series, such as Autoregressive (AR) and Vector Autoregressive (VAR) models, provide efficient baselines for one-step-ahead prediction. AR models predict future values of a single time series (e.g., HIP or NAc LFPs), while VAR models handle multiple correlated time series. However, their linear assumptions limit their ability to capture the nonlinear interactions inherent in neural data. Deep learning methods, like Long Short-Term Memory (LSTM) networks, offer greater flexibility but often rely solely on time-domain information, overlooking the complex oscillatory patterns in neural signals. Wavelet-based transformations, which provide time-frequency representations, have shown promise in isolating transient and periodic components critical to neural processing. Yet, applying these to multi-region brain data, particularly under varying reward conditions, remains challenging [3].

Wavelet coherence (WCOH) has emerged as a robust tool for quantifying instantaneous phase and amplitude relationships between two time series across frequency bands, revealing dynamic cross-spectral coupling. While widely used in other fields, its application to neurophysiological data for predictive modeling of dual-region activity is limited [4, 5]. This study introduces an enhanced prediction pipeline that integrates wavelet coherence with deep learning to forecast LFP signals from HIP and NAc under morphine, food, and saline conditions.

We employ a Wavelet Convolutional LSTM Stacked Autoencoder (WCLSA) for separate predictions of HIP and NAc LFPs, preserving region-specific dynamics. To capture cross-region connectivity, we introduce a novel Wavelet Coherence Convolutional LSTM Stacked Autoencoder (WCOH CLSA), which uses wavelet coherence to model nonlinear interactions between HIP and NAc. This dual approach enables both individual and joint predictions, offering a comprehensive view of reward processing across the three groups.

Our methodology transforms LFP signals into time-frequency space via Continuous Wavelet Transform (CWT). For WCLSA, this transformation supports independent predictions of HIP and NAc. For WCOH CLSA, wavelet coherence quantifies dynamic interdependence, expanding beyond linear models. These representations are fed into a stacked autoencoder with convolutional LSTM cells, extracting spatiotemporal features and projecting them onto a lower-dimensional manifold. Predictions are generated via Multi-Layer Perceptrons (MLPs)—separate for each region in WCLSA and shared for joint predictions in WCOH CLSA—preserving region-specific interpretability.

By comparing AR, VAR, LSTM, WCLSA, and WCOH CLSA across morphine, food, and saline groups, we aim to elucidate whether natural and pharmacological rewards engage similar or distinct neurocircuits, addressing a key debate in addiction research. Our findings may also inform the development of closed-loop neuromodulation systems requiring real-time brain state predictions. Overall, our results demonstrate the advantages of combining wavelet coherence with advanced deep learning architectures, while clarifying the distinct contributions of HIP and NAc during different reward conditions.

While prior studies have applied wavelet-based methods and deep learning to various domains, our contributions are novel in several ways. We are the first to apply WCLSA for separate predictions of HIP and NAc LFPs and to introduce WCOH CLSA for joint predictions, capturing cross-region connectivity in reward circuitry, and by combining wavelet coherence with a Convolutional LSTM Stacked Autoencoder, we model nonlinear interactions between signals, advancing beyond traditional linear models. Our approach maintains region-specific interpretability while modeling cross-region dynamics, a contrast to previous single-region or less interpretable multi-region models, and we systematically compare predictive performance across morphine, food, and saline groups, providing new insights into reward modulation of HIP–NAc connectivity. Our models offer a pathway for real-time interventions in neurological disorders, extending predictive modeling to clinical neuroscience, and we benchmark our models against AR, VAR, and LSTM, addressing a gap in multi-model evaluations for reward-related neural dynamics. This study advances brain signal prediction by introducing wavelet coherence-enhanced deep learning pipelines tailored to multi-region neural dynamics in reward circuitry, and by integrating WCLSA and WCOH CLSA, focusing on dual-region dynamics, and comparing across reward conditions, we enhance understanding of HIP–NAc connectivity and pave the way for innovative neuromodulation strategies

## 2. Methodology

This study compares multiple prediction models for LFPs recorded from the HIP and NAc in rats under three experimental conditions: morphine, food, and saline. The primary objective is to forecast one-step-ahead neural activity—both within each region independently and across the two regions jointly—so as to elucidate reward-related neural dynamics. Four main modeling approaches are examined:

1. **Autoregressive (AR) model** – a linear univariate baseline.
2. **Vector Autoregressive (VAR) model** – a linear multivariate extension for simultaneous prediction of HIP and NAc.
3. **Long Short-Term Memory (LSTM) model** – a deep learning, time-domain-only approach for univariate prediction.
4. **Wavelet Convolutional LSTM Stacked Autoencoder (WCLSA)** – a deep, nonlinear approach integrating wavelet-based time-frequency representations and convolutional LSTMs for *separate* HIP and NAc predictions.
5. Additionally, we introduce a novel **Wavelet Coherence Convolutional LSTM Stacked Autoencoder (WCOH CLSA)** that leverages wavelet coherence to capture instantaneous cross-region connectivity between HIP and NAc, thus providing *joint* predictions. This dual approach (WCLSA vs. WCOH CLSA) yields a comprehensive understanding of both local and inter-regional neural activity.

Below, we detail each methodological step: (1) data windowing, (2) time-frequency transformation, (3) wavelet coherence computation, (4) network architectures, (5) baseline model formulations, and (6) model selection and training.

1. Data Windowing

We apply a sliding window function to the raw LFP signals, generating overlapping segments that serve as training samples for one-step-ahead prediction. As shown in Fig. 1 (parts A and B), a window of length 12 samples is moved through the time series with an overlap of 11 samples (equivalent to a stride of 1). Formally, for a univariate LFP signal x(t), consecutive windows {x(t), x(t+1),…,x(t+11)} serve as the inputs for one-step-ahead prediction of x(t+12). When modeling both HIP and NAc simultaneously (e.g., in VAR or WCOH CLSA), the same windowing process is applied in parallel to each signal or to coherence features derived from the two signals.

2. Time-Frequency Transformation via Continuous Wavelet Transform (CWT)

To capture the oscillatory and transient features of neural data, each windowed segment is transformed into a time-frequency representation using the Continuous Wavelet Transform (CWT). Given a real-valued signal x(t), its wavelet transform $W_x(a,b;\psi)$ is computed by convolving x(t) with scaled and shifted versions of a mother wavelet $\psi(t)$ [6]. Mathematically:

$$W_x(a, b; \psi) = \frac{1}{\sqrt{a}} \int x(t)\psi^*\left(\frac{t-b}{a}\right) dt \tag{1}$$

where a is the scale factor, b is the shift factor, and $\psi(\cdot)$ is the mother wavelet. In this study, we use the Morlet wavelet (Fig. 1, part C) for its ability to capture both low-frequency and high-frequency information effectively [7, 8]. The resulting scalogram, or time-frequency map, is shown in Fig. 1, part D. For **WCLSA**, we generate this time-frequency representation separately for HIP and NAc, feeding each scalogram into its own prediction pipeline.

3. Wavelet Coherence (WCOH) Computation

In addition to single-region wavelet transforms, we compute **wavelet coherence** (WCOH) between HIP and NAc to characterize how their oscillatory activity co-evolves over time and across frequencies. Unlike standard CWT that focuses on a single signal, wavelet coherence measures localized phase and amplitude relationships between two signals x(t) (HIP) and y(t) (NAc). Formally, wavelet coherence at scale a and time b can be viewed as:

$$\text{WCOH}_{x,y}(a, b) = \frac{\left|\mathcal{S}[W_x(a,b)W_y^*(a,b)]\right|}{\mathcal{S}[|W_x(a,b)|^2]\mathcal{S}\left[|W_y(a,b)|^2\right]} \tag{2}$$

Where $W_x$ and $W_y$ are the wavelet transforms of x and y, respectively, and $\mathcal{S}[\cdot]$ denotes a smoothing operator across time and scale. Wavelet coherence maps (2D images) reveal frequency-specific coupling across time [9]. These maps are then used as inputs to the **WCOH CLSA** framework, allowing the network to learn dynamic cross-spectral interactions between HIP and NAc.

4. Deep Learning Architectures

Both **WCLSA** (for separate region predictions) and **WCOH CLSA** (for joint region predictions) share a similar architectural backbone, comprising:

1. **Convolutional LSTM cells** arranged in a stacked autoencoder configuration.
2. **Flattening layers** to reduce the extracted spatiotemporal features.
3. **Multi-Layer Perceptron (MLP) output layers** for one-step-ahead prediction.

4.1 Convolutional LSTM

The LSTM is a variant of recurrent neural networks (RNNs) designed to handle long-term dependencies via gating mechanisms. Convolutional LSTMs extend traditional LSTMs by replacing fully connected transformations with 2D convolutions, making them especially suitable for image-like inputs (e.g., wavelet scalograms) [10].

Let $X_t$ be the current input (a time-frequency map at time t), $H_{t-1}$ the hidden state, and $C_{t-1}$ the cell state. The convolutional LSTM operations are written as:

$$\begin{aligned}
i_t &= \sigma(W_{xi} * X_t + W_{hi} * H_{t-1} + W_{ci} \circ C_{t-1} + b_i), \\
f_t &= \sigma(W_{xf} * X_t + W_{hf} * H_{t-1} + W_{cf} \circ C_{t-1} + b_f), \\
\tilde{C}_t &= \tanh(W_{xc} * X_t + W_{hc} * H_{t-1} + b_c), \\
C_t &= f_t \circ C_{t-1} + i_t \circ \tilde{C}_t, \\
o_t &= \sigma(W_{xo} * X_t + W_{ho} * H_{t-1} + W_{co} \circ C_t + b_o), \\
H_t &= o_t \circ \tanh(C_t),
\end{aligned} \quad (3)$$

where $\sigma(\cdot)$ is the sigmoid function, $*$ denotes convolution, and $\circ$ is the Hadamard (elementwise) product [11]. By using convolutional filters, the network learns spatial (frequency-scale) correlations in wavelet-derived images while preserving temporal dependencies via the recurrent LSTM pipeline [12].

4.2 Stacked Autoencoder with Convolutional LSTMs

An autoencoder has two main components:

1. **Encoder** – compresses high-dimensional input into a lower-dimensional (latent) space.
2. **Decoder** – reconstructs the original input from the latent representation.

A *stacked* autoencoder consists of multiple encoder–decoder pairs arranged in succession, with the output (latent code) of one layer serving as the input to the next. In our **WCLSA** and **WCOH CLSA** frameworks, we replace the usual dense layers in the encoder/decoder blocks with convolutional LSTM cells [13, 14]. This design captures spatiotemporal features in wavelet (or wavelet coherence) maps, producing a compact latent representation well-suited for neural signal prediction.

Within each encoder block:

1. The input wavelet or wavelet coherence image is passed to a convolutional LSTM layer.
2. The hidden state is propagated to the next layer (or to the decoder if at the final encoder block).

## 4.3 MLP Output Layers

After feature extraction via the stacked convolutional LSTM autoencoder, we apply **flattening** layers to transform the final 3D (height × width × channel) feature maps into 1D vectors. These vectors are passed to a two-layer MLP (multi-layer perceptron) to predict the next sample. As shown in Fig. 1 (top panel, red layers), the MLP effectively acts as the regression head on top of the learned spatiotemporal features.

- **WCLSA** produces separate predictions for HIP and NAc by training two identical pipelines, each receiving wavelet scalograms of the respective region.
- **WCOH CLSA** incorporates wavelet coherence maps (between HIP and NAc) into the convolutional LSTM autoencoder, enabling the model to learn dynamic interdependencies across regions. Two MLP output head for separate predictions (one for each region).

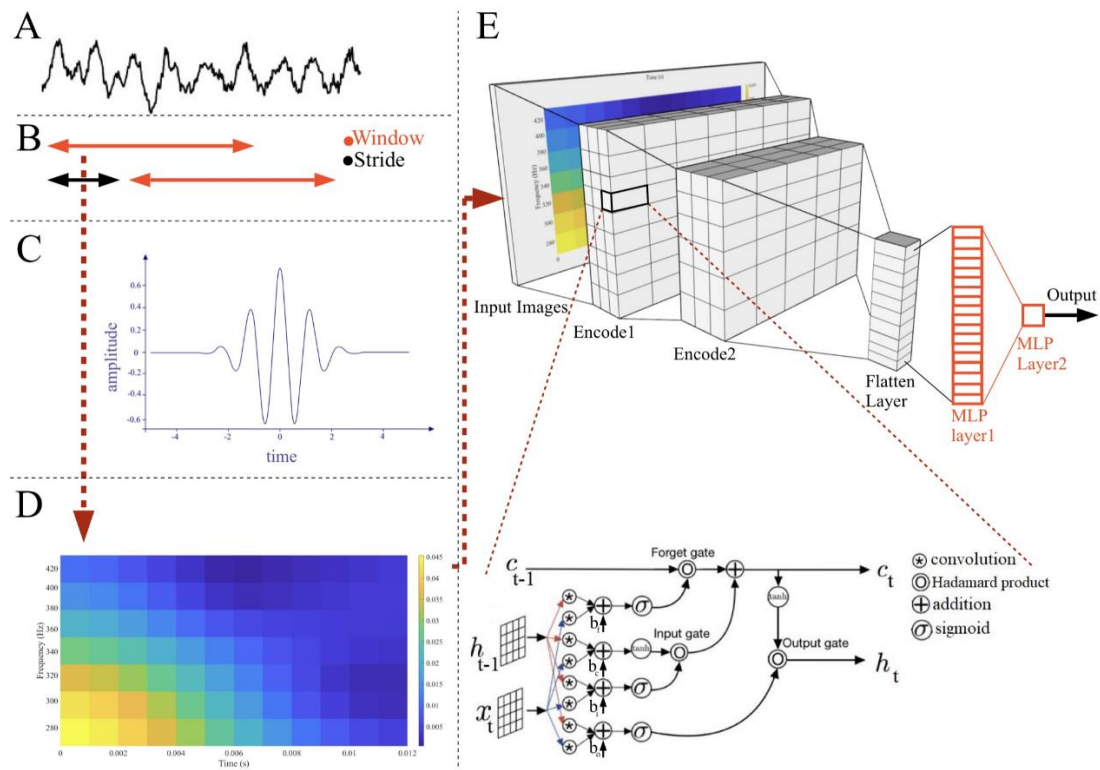

Fig. 1. WCLSA (WCOH CLSA) structure: A. Time-domain input signal, B. Data sampling structure: sampling window length and sampling stride, C. Morlet wavelet, D. Time-frequency representation of input data using wavelet\WCOH transforms, E. Top panel: Structure of one-step

prediction using two layers MLP (red color) following stacked autoencoder with convolution LSTM cells, Bottom panel: convolution LSTM structure.

## 5. Comparative Models

To benchmark our deep learning frameworks, we compare them with traditional linear models (AR and VAR) and a baseline deep learning method (LSTM).

### 5.1 AR Model (Univariate)

An autoregressive model of order pp predicts a time series $Y_t$ using its own past values:

$$Y_t = \beta_0 + \beta_1 Y_{t-1} + \beta_2 Y_{t-2} + \cdots + \beta_p Y_{t-p} + \epsilon_t \qquad (4)$$

where $\epsilon_t$ is white noise and $\beta_i$ are model coefficients. We fit separate AR models for HIP and NAc signals independently.

### 5.2 VAR Model (Multivariate)

Vector autoregression (VAR) extends AR to handle multiple, potentially correlated time series. For two signals $Y_1$ (HIP) and $Y_2$ (NAc), a VAR define as [15]:

$$\begin{aligned} Y_{1,t} &= \alpha_1 + \beta_{11,1} Y_{1,t-1} + \beta_{12,1} Y_{2,t-1} + \epsilon_{1,t} \\ Y_{2,t} &= \alpha_2 + \beta_{21,1} Y_{1,t-1} + \beta_{22,1} Y_{2,t-1} + \epsilon_{2,t}. \end{aligned} \qquad (5)$$

For higher orders, additional lags are included. We determined the optimal lag (12 samples) for our LFP data using the Akaike Information Criterion (AIC) and Bayesian Information Criterion (BIC). VAR thus captures linear cross-dependencies between HIP and NAc but cannot model the nonlinear couplings that our wavelet-based deep learning approaches exploit [16].

### 5.3 Baseline LSTM (Time-Domain Only)

To highlight the effect of wavelet transformations, we also include a simple **LSTM** baseline that operates purely in the time domain. Each windowed signal segment $\{x(t), \ldots, x(t+11)\}$ directly feeds into the LSTM layer(s), with a subsequent dense layer producing the one-step-ahead prediction. This comparison underscores the potential advantage of extracting time-frequency features prior to LSTM-based modeling.

## 6. Model Selection, Training, and Evaluation

All models (AR, VAR, LSTM, WCLSA, WCOH CLSA) were trained separately on LFP data from the three groups: morphine, saline, and food. To mitigate overfitting, we used a cross-validation approach (10-fold) where each model's hyperparameters and performance were assessed on multiple partitions of the data. Predictive accuracy was quantified as the coefficient of determination (R²). Optimal hyperparameters (e.g., learning rate, number of convolutional LSTM filters, MLP layer sizes) were selected via grid search or Bayesian optimization, using a validation set for early stopping.

- **Training Loss and Optimization**: Mean Squared Error (MSE) was used as the loss function for deep models, minimized via backpropagation with an Adam optimizer.
- **Regularization**: Techniques such as dropout, weight decay, or partial reconstruction (in the stacked autoencoder) mitigated overfitting.
- **Model Selection**: Performance on a held-out test set was evaluated using MSE. The best model for each data group was determined by lowest prediction error and/or highest explained variance.

7. Summary of Methodological Contributions

1. **Windowing and Wavelet Transforms**: We transform short temporal windows of LFP signals into detailed time-frequency maps using Morlet-based CWT.
2. **Wavelet Coherence Analysis**: We extend single-region wavelet transforms to wavelet coherence for capturing HIP–NAc cross-spectral interactions.
3. **Deep Autoencoder with Convolutional LSTMs**: Convolutional LSTMs replace conventional dense layers in a stacked autoencoder, effectively encoding and decoding spatiotemporal features from wavelet (and wavelet coherence) images.
4. **One-Step Prediction with MLP**: A two-layer MLP processes the latent vectors to forecast the next sample, thereby fusing deep feature extraction with a flexible regression head.
5. **Comparison with Linear and LSTM Baselines**: AR, VAR, and standard LSTM models highlight the benefits of time-frequency representations and convolutional LSTM architectures.
6. **Reward Condition Insights**: By separately modeling morphine-, food-, and saline-treated groups, we identify how pharmacological vs. natural rewards modulate intra- and inter-regional neural dynamics.

This integrated methodology, especially the novel **WCOH CLSA** pipeline, not only enables accurate one-step-ahead predictions in LFP data but also provides a lens into the underlying connectivity of neural circuits in the reward system. By contrasting the performance and interpretability of linear, time-domain-only, and advanced wavelet-based deep learning models, we aim to clarify how natural and pharmacological rewards differentially engage HIP–NAc circuitry.

## 3. RECORDED DATA

1. Animals and Surgery

At the beginning of each experiment, a male Wistar rat (Pasteur Institute, Tehran, Iran) weighing 220–270 g was maintained on a controlled condition (12/12 h light/dark cycle in temperature (25 ± two °C) and humidity (55±10%). Before the conditioning phase, rats were fed 80-85% of their free-feeding body weight [15]. In accordance with the National Institutes of Health (NIH) guidelines (NIH publication 80-23 revised in 1996), the Ethics Committee approved all studies conducted at the Shahid Beheshti University of Medical Sciences (IR.SBMU.SM.REC.1395.373) in Tehran, Iran. The rats were anesthetized [17] and implanted with the bipolar recording electrodes aiming to record local field potentials (LFPs) in the hippocampal CA1 [18] as well as NAc at the following coordinates, respectively: anteroposterior (AP): -3.4 mm from bregma, lateral (L): ±2.5 mm, dorsal-ventral (DV): -2.6 mm and for NAc: AP: 1.5 mm, L: ±1.5 mm, DV: -7.6 mm. The reference and ground screws were inserted into the skull. Rats recovered for one week after surgery. Using a rat brain atlas, the electrode tip traces were localized and confirmed at the end of the experiments [19].

2. Conditioned place preference paradigm

All rats were exposed to an unbiased, counterbalanced conditioned place preference (CPP) procedure LFP recordings data were performed simultaneously from hippocampal CA1 and NAc in free-moving animals in a pre-and post-test of CPP. The CPP paradigm includes three phases: pre-conditioning, conditioning, and post-conditioning. The experiments were performed in a three-compartment Plexiglas CPP apparatus, consisting of two equal-sized compartments as the main chambers for conditioning reward and a smaller chamber (Null) connecting the two main chambers. The floor texture (smooth or rough) and wall stripes pattern made the two main compartments different. There is a horizontal stripe on one compartment's wall and a vertical stripe on the other compartment's wall [20].

Animals' behavior was monitored using a 3CCD camera (Panasonic, Japan) that was positioned above the apparatus. Data were analyzed using the Ethovision software (Noldus Information Technology, the Netherlands), a video tracking system for automating behavioral experiments that were programmed to simultaneously trigger the onset of behavioral tracking and the beginning of LFPs recording. Therefore, behavior and electrophysiological sessions were recorded in a sync manner.

During the pre-and post-test, rats freely explored the entire arena for 10 minutes while they were connected to the LFP recording cable. Rats that showed an inherent preference >80% for the main compartment of the CPP were removed from the experiment. The distance traveled and time spent in each of the compartments were recorded [21].

3. Conditioning phase (Saline, Morphine, Food)

On the first day of the conditioning phase, each animal received morphine (5 mg/kg, s.c.) in the morning and was confined to one main chamber of the CPP compartment for 30 min; about six hours later, they received saline (1 ml/kg, s.c.) and were confined to another main chamber for 30 min. The third day of conditioning was the same as the first day. During this phase, access to other chambers of the CPP box was blocked. In the natural (food) group, on the first day of the conditioning period, in the morning session, food-restricted animals received 6 g biscuit as a reward in the middle of the one main compartment, and six hours later, they were placed into the other compartment with no food; each session lasted 30 min. On the following days, biscuit and no-food session times were arranged in a counterbalanced manner over the conditioning period [22].

4. Post-conditioning phase (Post-test)

Twenty-four hours after the conditioning phase, rats were tested for a post-test trial (10-min) in which they could explore the entire CPP arena, like the pre-test. The behavioral and LFP data were measured during this session.

5. Electrophysiological and behavioral recordings

Behavior was recorded with a digital video camera (30 frames per second), and rat movements were tracked by an automated system that synchronized behavioral data and electrophysiological recordings. In each frame, the spatial position is defined as the center of the animal body. During the experiments, a lightweight, flexible cable was connected to the pins on the preamplifier. Recordings, digitalization, and filtering of neural activities were performed using a commercial acquisition processor.

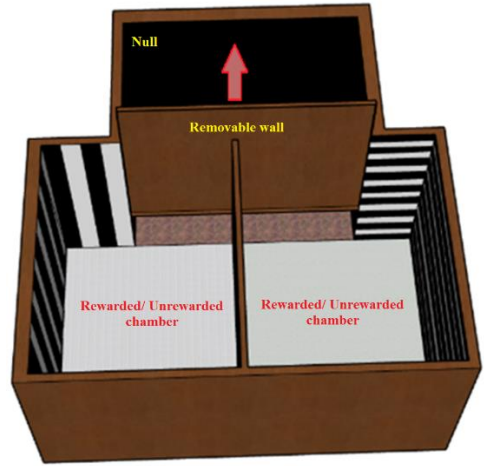

Fig. 2. Schematic of the CPP box

## 4. Results

This study examined four predictive modeling approaches for one-step-ahead prediction of Local Field Potentials (LFPs) recorded from the Hippocampus (HIP) and Nucleus Accumbens (NAc) in rat brains across three experimental groups: morphine, food, and saline. The evaluated models included:
1. **Autoregressive (AR)**: A linear method predicting HIP and NAc activities individually.
2. **Vector Autoregression (VAR)**: A multivariate linear model predicting HIP and NAc simultaneously.
3. **Long Short-Term Memory (LSTM)**: A simpler nonlinear neural network.
4. **Wavelet Convolution LSTM Stacked Autoencoder (WCLSA)**: An advanced deep nonlinear model integrating wavelet transforms and convolutional LSTM architecture.
5. **Wavelet coherence LSTM Stacked Autoencoder (WCOH CLSA)** was employed to analyze HIP-NAc connectivity and its influence on predictive performance, providing insights into both linear and nonlinear interactions across time-frequency domains.

Figure 3 provides a schematic overview of these methodologies.

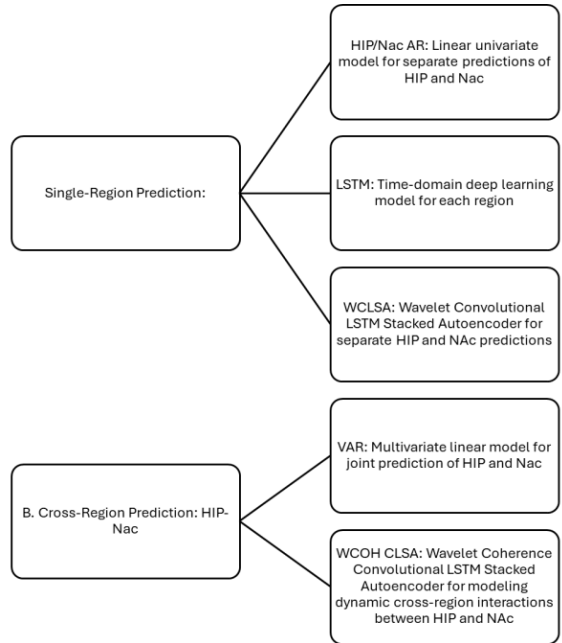

Figure 3. Predictions method

1. Correlation and Connectivity between HIP and NAc

The effectiveness of the VAR model relies on the correlation between HIP and NAc signals, making it less suitable when such correlation is weak or absent. To assess this, we computed the average linear correlation coefficients between HIP and NAc LFPs for each experimental group, as presented in **Table 1**.

**Table 1: Average HIP-NAc Linear Correlation**

| Group | Correlation |
|---|---|
| Morphine | -0.06 |
| Food | 0.40 |
| Saline | 0.02 |

The food group exhibited a significant positive correlation (0.40), indicating coordinated activity between HIP and NAc. In contrast, the morphine group showed a slight negative correlation (-0.06), and the saline group displayed near-zero correlation (0.02), indicating no significant connectivity between HIP and NAc, which is associated with the absence of a learning process in this neutral condition. These findings suggest that the VAR model is primarily applicable to the food group, where inter-regional relationships enhance its predictive capability.

To further explore connectivity, WCOH was analyzed to capture both linear and nonlinear interactions between HIP and NAc across time-frequency domains. Unlike linear correlation, WCOH provides a dynamic measure of phase synchronization, revealing the strength and frequency-specific nature of HIP-NAc coupling. The WCOH values were integrated into the predictive accuracy assessments, offering a unique perspective on how connectivity influences model performance. Compared to linear correlation, WCOH provides a more nuanced understanding of connectivity, capturing nonlinear dynamics that linear measures may miss. This distinction is particularly evident in the morphine group, where the negative correlation (-0.06) contrasts with WCOH's ability to detect residual nonlinear coupling, and in the food group, where WCOH aligns closely with the positive correlation (0.40) but offers additional frequency-specific insights.

2. Predictive Model Performance and Comparison

All five models—AR, VAR, WCLSA, WCOH CLSA and LSTM—were applied to LFP signals from HIP and NAc across the morphine, food, and saline groups. Predictive accuracy was evaluated as the average performance across rats in each condition.
**Table 2** and **Table 3** summarize the predictive accuracies for NAc and HIP, respectively.

**Table 2: Average Predictive Accuracy for NAc**

| Model | Morphine | Food | Saline |
|---|---|---|---|
| AR | 0.75 | 0.71 | 0.73 |
| LSTM | 0.83 | 0.80 | 0.82 |
| WCLSA | 0.96 | 0.94 | 0.92 |
| VAR | 0.48 | 0.89 | 0.55 |
| WCOH | 0.75 | 0.92 | 0.60 |

**Table 3: Average Predictive Accuracy for HIP**

| Model | Morphine | Food | Saline |
|---|---|---|---|
| AR | 0.72 | 0.76 | 0.73 |
| LSTM | 0.81 | 0.85 | 0.84 |
| WCLSA | 0.93 | 0.97 | 0.92 |
| VAR | 0.46 | 0.91 | 0.57 |
| WCOH | 0.70 | 0.97 | 0.59 |

The WCLSA model consistently outperformed all other methods across both regions and all experimental conditions, achieving accuracies up to 0.97 (HIP, food group) and 0.96 (NAc, morphine group) and 0.92 (NAc, saline group). This superior performance highlights the advantage of integrating time-frequency analysis (via Continuous Wavelet Transform) with deep learning architectures for capturing the complex dynamics of LFP signals. The LSTM model ranked second in the morphine and saline groups, followed by the AR model, while the VAR model excelled in the food group (NAc: 0.89, HIP: 0.91), surpassing AR and LSTM due to the positive HIP-NAc correlation. Given VAR's lower computational complexity compared to WCLSA, it offers a practical alternative in contexts where inter-regional connectivity is significant.
WCOH provided unique insights into the connectivity dynamics and their impact on prediction. In the food group, WCOH yielded high predictive accuracies (NAc: 0.92, HIP: 0.97), closely aligning with WCLSA (NAc: 0.94, HIP: 0.97) and surpassing VAR (NAc: 0.89, HIP: 0.91). This suggests that WCOH effectively captures the strong, possibly nonlinear, HIP-NAc coupling in natural reward contexts, enhancing predictive accuracy beyond linear correlation alone. Compared to VAR, WCOH offers enhanced sensitivity to frequency-specific coupling, making it particularly valuable for understanding natural reward processing. In contrast, in the morphine group, WCOH accuracies were lower (NAc: 0.75, HIP: 0.70) compared to WCLSA (NAc: 0.96, HIP: 0.93) and diverged from the VAR model (NAc: 0.48, HIP: 0.46). This indicates that while WCOH captures some connectivity, the relationship may be nonlinear or negatively correlated, as suggested by the negative linear correlation (-0.06). Compared to WCLSA, WCOH's moderate performance in the morphine group underscores WCLSA's ability to model complex, region-specific dynamics without relying on inter-regional coherence. For the saline group, WCOH accuracies (NAc: 0.60, HIP: 0.59) were moderate, reflecting minimal connectivity influence under neutral conditions, consistent with the near-zero correlation (0.02).

3. Differential Accuracy and Connectivity Insights

A comparative analysis of predictive accuracies between HIP and NAc across groups revealed nuanced differences:
- **Food Group**: HIP accuracy exceeded NAc accuracy across models (WCLSA: 0.97 vs. 0.94; WCOH: 0.97 vs. 0.92; VAR: 0.91 vs. 0.89), indicating that HIP activity is more predictable and potentially more central to natural reward processing. The high WCOH values and positive correlation (0.40) suggest enhanced HIP-NAc connectivity, supporting coordinated network activity in food reward contexts. This connectivity likely contributes to the superior performance of VAR and WCOH in this group, with WCOH offering additional insights into frequency-specific coupling compared to VAR.
- **Morphine Group**: NAc accuracy surpassed HIP accuracy across models (WCLSA: 0.96 vs. 0.93; WCOH: 0.75 vs. 0.70; VAR: 0.48 vs. 0.46), underscoring NAc's greater predictability and likely dominance in opioid-induced reward processing. The lower WCOH values and negative correlation (-0.06) suggest a disrupted or nonlinear HIP-NAc relationship, which may reflect morphine's pharmacological disruption of natural reward circuitry. The poor performance of VAR in this group further confirms that linear connectivity does not aid prediction, while WCOH captures some residual nonlinear coupling, though its predictive utility is limited compared to WCLSA's region-specific modeling.
- **Saline Group**: No significant difference was observed between NAc and HIP accuracies across models (WCLSA: 0.92 vs. 0.92; WCOH: 0.60 vs. 0.59; VAR: 0.55 vs. 0.57), consistent with the near-zero correlation (0.02), indicating negligible connectivity influence under neutral conditions. Here, WCOH provides slightly better than, capturing minimal connectivity that linear models miss, though its predictive accuracy remains moderate.

These findings highlight the differential engagement of HIP and NAc in distinct reward types. The superior performance of HIP-NAc connectivity measures (VAR and WCOH) in the food group suggests a tightly coupled network, with WCOH offering enhanced sensitivity to nonlinear dynamics compared to VAR. In contrast, in the morphine group, WCOH indicates a nonlinear or negative relationship, potentially missed by linear correlation alone, though its predictive utility is limited compared to WCLSA. This underscores the importance of considering both linear and nonlinear connectivity measures in understanding brain network dynamics.

4. Conclusion

This study provides a comprehensive analysis of predictive modeling of Local Field Potentials (LFPs) in the rat brain, focusing on the Hippocampus (HIP) and Nucleus Accumbens (NAc) within the reward circuitry. By benchmarking five models—Autoregressive (AR), Vector Autoregression (VAR), Long Short-Term Memory (LSTM), Wavelet Convolution LSTM Stacked Autoencoder (WCLSA)—and incorporating wavelet coherence Convolution LSTM Stacked Autoencoder (WCOH CLSA) analysis, we elucidated the predictive accuracy and connectivity dynamics across morphine, food, and saline conditions.

1. Methodological Contributions

The development of the WCLSA model represents a significant advancement, achieving the highest predictive accuracy across all groups (up to 0.97 for HIP in food, 0.96 for NAc in morphine and 0.92 in saline). By leveraging Continuous Wavelet Transform (CWT) for time-frequency transformation and convolutional LSTM cells within a stacked autoencoder, WCLSA excels at modeling the nonlinear and nonstationary nature of LFP signals. The integration of wavelet coherence (WCOH) further enriches this framework by capturing inter-regional connectivity, offering a novel dimension to predictive neuroscience. Notably, WCOH provided a unique lens to assess how connectivity influences predictive performance, revealing its strengths in natural reward contexts and limitations in pharmacologically altered states. Compared to VAR, WCOH offers enhanced sensitivity to frequency-specific coupling, making it particularly valuable for understanding natural reward processing, while WCLSA's superior performance across all conditions underscores its versatility in handling both connected and disconnected network states.

2. Analytical Insights

The WCLSA model's dominance underscores the value of deep nonlinear approaches, yet the VAR model's strong performance in the food group (NAc: 0.89, HIP: 0.91) highlights the utility of simpler multivariate models when HIP and NAc are correlated. Wavelet coherence analysis revealed that HIP-NAc coupling is particularly robust in the food group (WCOH: 0.92–0.97), supporting the VAR findings and demonstrating WCOH's ability to capture both linear and nonlinear connectivity. Compared to VAR, WCOH offers additional insights into frequency-specific coupling, making it particularly valuable for natural reward processing. In contrast, in the morphine group, lower WCOH values (0.70–0.75) and a negative correlation suggest a complex, possibly nonlinear relationship that WCOH partially captures but does not fully leverage for prediction, unlike WCLSA's region-specific modeling. Compared to WCLSA, WCOH's moderate performance in the morphine group underscores WCLSA's ability to model complex, region-specific dynamics without relying on inter-regional coherence.

3. Neuroscientific Implications

The differential predictive accuracies illuminate the specialized roles of HIP and NAc in reward processing. HIP's higher accuracy in the food group (0.97 vs. 0.94) and enhanced connectivity with NAc point to its pivotal role in natural reward encoding, consistent with its involvement in memory and contextual learning. Conversely, NAc's superior accuracy in the morphine group (0.96 vs. 0.93) aligns with its established role in mediating pharmacological reward and addiction [23]. In the saline group, the predictive accuracies for both NAc and HIP using WCLSA are lower (both 0.92) compared to the higher accuracies observed in the morphine group for NAc (0.96) and in the food group for HIP (0.97). This lower performance, also evident in connectivity-based models (e.g., VAR: 0.55-0.57, WCOH: 0.59-0.60), indicates a baseline neural state where, in the absence of reward or pharmacological stimuli, no learning process occurs to enhance functional connectivity between HIP and NAc. These findings suggest that reward type modulates the functional interplay between HIP and NAc, with natural rewards fostering integration and pharmacological rewards potentially disrupting it. The integration of WCOH further refines this understanding by quantifying connectivity strength and nature, with WCOH offering enhanced sensitivity to nonlinear dynamics compared to VAR.

## 4. Future Directions

While this study establishes a robust predictive framework, further investigation into HIP-NAc connectivity is warranted. Advanced techniques such as phase-amplitude coupling could unravel the directional and oscillatory dynamics underlying these interactions, particularly the nonlinear or negative relationships hinted at by WCOH in the morphine group. Additionally, exploring how these connectivity patterns evolve over time or in response to different stimuli could provide deeper insights into reward circuitry plasticity.

## 5. Final Remarks

In conclusion, this work advances both the methodological and neuroscientific understanding of brain signal prediction in reward contexts. By demonstrating the efficacy of WCLSA, the context-specific utility of VAR, and the connectivity insights from wavelet coherence, it provides a versatile toolkit for studying neural dynamics. The differential roles of HIP and NAc across reward types, coupled with connectivity analyses, offer a quantitative foundation for future studies, with profound implications for addiction neuroscience and therapeutic development.